\documentclass[aps,showpacs,prb,superscriptaddress,floatfix,twocolumn,citeautoscript]{revtex4-1}
\usepackage{graphicx}
\usepackage{amsmath}
\usepackage{amssymb}
\usepackage{bm}
\usepackage{dcolumn}
\usepackage{subfigure}
\usepackage{units}
\usepackage{hyperref}
\usepackage[usenames]{color}
\usepackage{booktabs}
\usepackage{multirow}
\usepackage{siunitx}
\usepackage{wrapfig}
\usepackage{lipsum}
\usepackage{setspace}
\hypersetup{pdfborder=0 0 0,colorlinks=true,citecolor=blue,linkcolor=blue}
\input epsf

\newcommand{\MS}{MoS$_2$}

\newcommand{\ket}[1]{\left| #1 \right>} % for Dirac bras
\newcommand{\bra}[1]{\left< #1 \right|} % for Dirac kets
\begin{document}
	
	 \title{Strong anisotropic optical conductivity in two dimensional puckered structures: The role of Rashba effect}
	
	\author{S. Saberi-Pouya}
	\affiliation{Department of Physics, Shahid Beheshti University, G. C., Evin, Tehran 1983969411, Iran}
	\affiliation{Department of Physics, University of Antwerp, Groenenborgerlaan 171, B-2020 Antwerpen, Belgium}
	\author{T. Vazifehshenas}
	\email{t-vazifeh@sbu.ac.ir}
	\affiliation{Department of Physics, Shahid Beheshti University, G. C., Evin, Tehran 1983969411, Iran}
	\author{T. Salavati-fard}
	\affiliation{Department of Physics and Astronomy, University of Delaware, Newark, DE 19716, USA}
	\author{M. Farmanbar}
	\affiliation{Faculty of Science and Technology and MESA$^+$ Institute for Nanotechnology, University of Twente, P.O. Box 217, 7500 AE Enschede, The Netherlands}
	\author{F. M. Peeters}
	\affiliation{Department of Physics, University of Antwerp, Groenenborgerlaan 171, B-2020 Antwerpen, Belgium}
\begin{abstract}	
	
	We calculate the optical conductivity of an anisotropic two-dimensional system with Rashba spin-flip excitation within the Kubo formalism. We show that the anisotropic Rashba effect caused by an external field changes significantly the magnitude of the spin splitting. Furthermore, we obtain an analytical expression for the longitudinal optical conductivity associated with inter-band transitions as a function of the frequency for an arbitrary polarization angle. We find that the diagonal components of the optical conductivity tensor are direction-dependent and the spectrum of optical absorption is strongly anisotropic with an absorption window. The height and width of this absorption window are very sensitive to the system anisotropy. While the height of absorption peak increases with increasing effective mass anisotropy ratio, the peak intensity is larger when the light polarization is along the  armchair direction. Moreover, the absorption peak width becomes broader as the density of state mass or Rashba interaction is enhanced. These features can be used to determine parameters relevant for spintronics through the optical absorption spectrum.

\end{abstract}

	\date{\today}

	\maketitle

	\section{Introduction}
	
 Phosphorene, a monolayer of black phosphorus (BP), with puckered structure has attracted considerable attention because of the unique physical properties associated with its anisotropic band structure \cite{Phosphorene:review,Ling:pnas15,PhysRevB.94.085417,PhysRevB.95.045422,PhysRevB.95.041406}.	BP is a layered material in which each layer forms a puckered surface due to sp$^3$ hybridization. In its bulk crystalline form, BP is a semiconductor with a direct band gap of 0.3 eV which reaches up to 2 eV in the monolayer structure \cite{PhysRevB.89.235319,0957-4484-26-21-215205}.
Phosphorene-like materials, group IV-VI compounds, resemble in many respects, for instance, in-plane anisotropy, orthorhombic lattice and puckered layered structure\cite{0957-4484-27-27-274001,PhysRevB.92.085406,PhysRevB.93.125428}.
Similar to phosphorene and as a consequence of their orthorhombic structure, transport, optoelectronic spintronic properties of these materials are highly anisotropic \cite{Med:JPC16,2DSnS}.  
	
Nowadays, the spin-orbit coupling interaction is a field of great interest owing to potential applications in spintronic phenomena and electric manipulation of spins \cite{RevModPhys.76.323,PhysRevB.74.165310,SpintronicsNat:MacDonald,PhysRevLett.94.047204,PhysRevLett.112.086802,Spinquantumgas:2013,Alhili:jcg15}. This interaction shows up lacking a center of inversion symmetry  in crystalline lattices (the Dresselhaus type \cite{PhysRev.100.580}) or structural asymmetry in the interfaces/surface region of quantum wells (the Rashba type \cite{rashba1960properties}). In two dimensional (2D) materials, when the inversion symmetry is broken by an applied transverse electric field or a substrate, the spin degeneracy is lifted due to the Rashba effect \cite{KaneMele:prl2005, bychkov1984oscillatory, NewperspectivesRSOC, KaneMele:prl2005, PhysRevB.74.245309}. Therefore, transitions between spin split states results in a non-zero value for the optical conductivity in the presence of an alternating electric field. 
 The absorption part of the optical conductivity can be used in order to probe the spin-split energy levels. The spintronic parameters such as the Rashba coupling strength, electron density and also spin polarization in the 2D materials can be measured optically \cite{PhysRevB.72.033320, PhysRevB.74.075321, Ang2014,PhysRevB.94.155432, PhysRevB.86.205425, PhysRevLett.105.136805, PhysRevB.94.161404}.
 
 Recently, Xiao \cite{PhysRevB.94.155432} et al. studied the optical conductivity of \MS\ in the presence of spin-orbit coupling and found that the Rashba spin-orbit parameter can change the absorption peak or window in the optical spectrum. In contrast to the isotropic band structure of \MS\, the anisotropic band structure in phosphorene results in a highly anisotropic Rashba splitting, \cite{PhysRevB.94.155423,PhysRevB.92.035135} hence, the strength of spin splitting depends on the direction of $k$, as well as its magnitude \cite{PhysRevB.92.035135}. The anisotropic Rashba spin-orbit interaction in 2D electron or hole gas systems due to the k-cubic Rashba spin-orbit interaction gives rise to different features in the optical conductivity, anisotropic spin susceptibility and also distinctive behavior of the spin Hall conductivity \cite{PhysRevB.91.241302,0953-8984-28-42-425302}.  Another anisotropic behavior of the spin splitting appears for the interplay between both Rashba and Dresselhaus SOC in a 2D electron gas. It has been shown that the  anisotropic dynamical optical conductivity can be used as a powerful tool to probe and manipulate the coupling strengths and set out the range of frequency where the optical conductivity is essentially non-zero. \cite{PhysRevB.94.161404, PhysRevB.95.035401}

	\begin{figure*}
		\includegraphics[width=18.5cm]{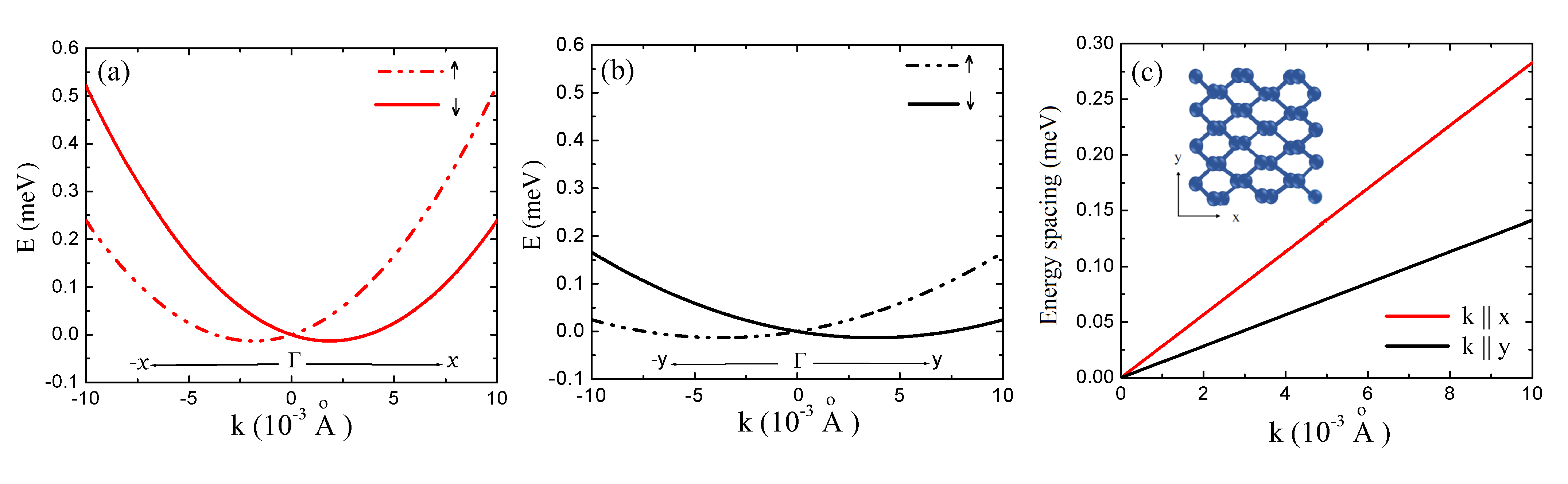}
		\caption{(a) Energy dispersion along $x$ direction ($\theta$=0) and (b) along $y$ direction ($\theta=\pi/2$) with $m_x$=$m_0$, $m_y$=4$m_0$, $\alpha_{R}$=10meV$\mathring{A}$ (c) spin-splitting along $x$ and $y$ direction for several values of $\alpha_{R}$ with  $m_x$=$m_0$  and $m_y$=4$m_0$. The inset figure is a top view of an anisotropic monolayer with the $x$ and $y$ axes being the armchair
			and zigzag directions, respectively.}
		\label{fig1}
	\end{figure*}
	
In this paper, we use the $k$-linear Rashba Hamiltonian for anisotropic 2D materials, such as phosphorene and group-IV monochalcogenides which have been predicted to exhibit an anisotropic energy band  \cite{liu:nano15,Xia:nat13,PhysRevB.92.085406,Med:JPC16}. We find that the extrinsic spin-orbit coupling due to broken inversion symmetry has a strong anisotropic nature which impacts the optical response of 2D electron systems. Subsequently, we calculate the optical conductivity of an anisotropic 2D material with paraboloidal energy band using the Kubo formalism and show that the absorption peak in the optical spectrum is very sensitive to the anisotropic effective mass ratio. The most significant contribution to the optical absorption occurs when the polarization of light is along the armchair direction (the direction with a smaller effective mass). Whereas phosphorene is a good example of an 2D anisotropic system, our formalism also applicable to other 2D puckered materials.

Our paper is organized as follows. In Section \ref{theory1}, we describe the basic Hamiltonian used in this work in the presence of anisotropic Rashba effect and develop a general formalism applicable for anisotropic 2D materials. We obtain in Section \ref{theory2} an analytical expression for the optical conductivity due to spin-flip transitions. We present our numerical results and provide a discussion of our findings. The highlights of this work are summarized in Section \ref{conclusion}.
		
	\begin{figure*}
		\includegraphics[width=18.0cm]{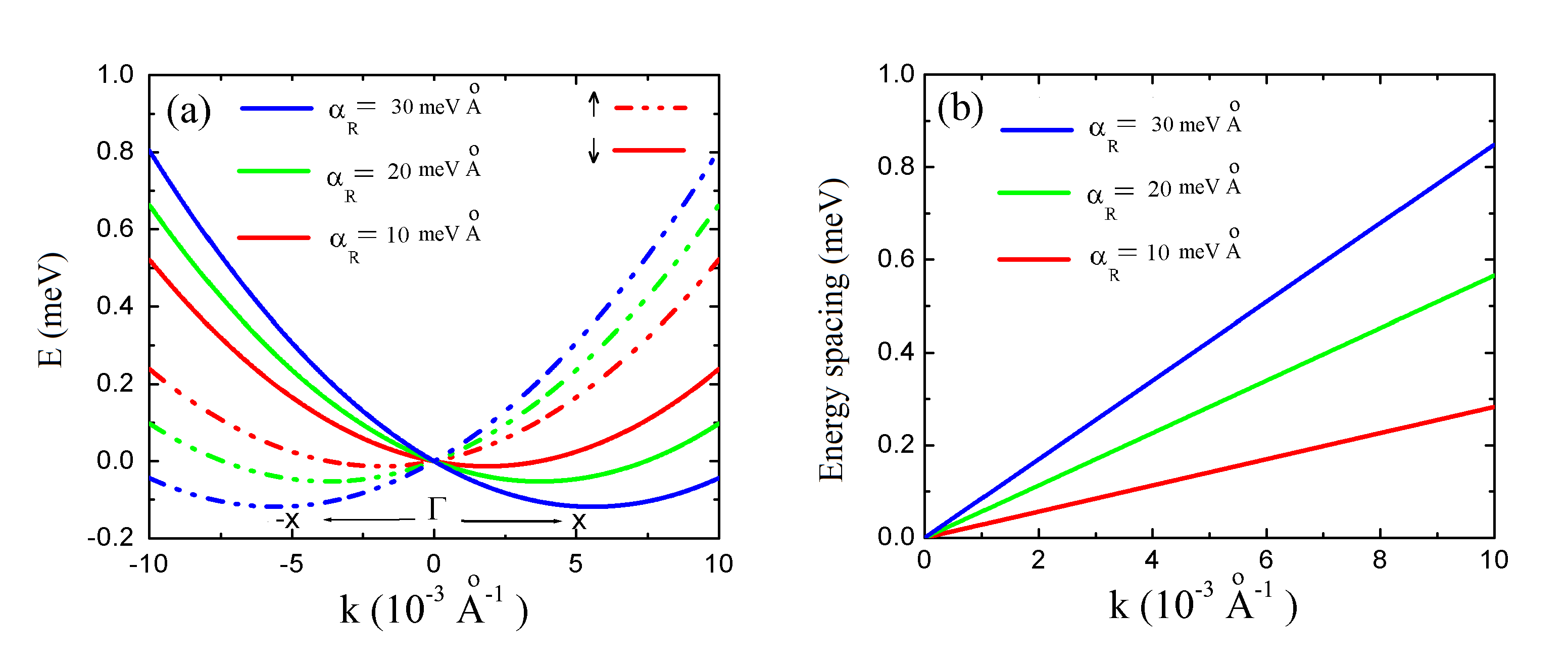}
		\caption{(a) Energy dispersion and (b) spin-splitting along $ x $ (armchair) direction for a few values of $\alpha_{R}$ with $m_y$=4$m_0$ and $m_x$=$m_0$}
		\label{fig2}
		\end{figure*}

\section{Paraboloidal Energy Band with Anisotropic Rashba Effect} \label{theory1}

  We study the low energy dispersion and optical conductivity of a 2D system with paraboloidal energy band in the presence of Rashba spin-orbit coupling. We assume that the 2D system is a puckered honeycomb lattice where the $x$ and $y$ axes are taken to be along its armchair and zigzag directions, respectively. The Hamiltonian for such a system including the extrinsic Rashba term is given by:
 
 \begin{equation}
 \begin{aligned}
 \hat{H}=\hat{H}_{0}+\hat{H}_{R} \ .    
 \label{eq:1}
 \end{aligned}
 \end{equation} 
\noindent Here $\hat{H}_{0}$ is the $\mathbf{k}.\mathbf{p}$ free electron Hamiltonian whose low energy spectrum for an anisotropic 2D system is obtained as\cite{Rodin:prl2014}: 

\begin{equation}
\hat{H}_{0}=\frac{\hbar^2}{2}(\frac{k_{x}^{2}}{m_{x}}+\frac{k_{y}^{2}}{m_{y}})\hat{\sigma_{0}} \ ,
\label{eq:3}
\end{equation}

\noindent where  $m_{x}$ and $m_{y}$ are the electron effective masses along $x$ and $y$ directions respectively, and $\hat{\sigma_{0}}$ is a $2\times2$ unitary matrix.
The Rashba anisotropic Hamiltonian which has been recently introduced for phosphorene \cite{PhysRevB.92.035135} can be rewritten as:

\begin{equation}
\begin{aligned}
\hat{H}_{R}=-\alpha_{R} (\sqrt{\frac{m_d}{m_x}}k_x \hat{\sigma_y}-\sqrt{\frac{m_d}{m_y}}k_y \hat{\sigma_x}) \ .    
\label{eq:4}
\end{aligned}
\end{equation}

\noindent Here, $\alpha_{R}$ is the Rashba coefficient, $m_d=\sqrt{m_x m_y}$ is the density of state mass and $\hat{\sigma_x}$ and $\hat{\sigma_y}$ are the Pauli matrices. 

Upon diagonalizing the total Hamiltonian, one obtains two branches of the energy spectrum:

\begin{equation}
\begin{aligned}
E^{\lambda}(\mathbf{k})=\frac{\hbar^{2} k^{2} R(\theta)}{2}+\lambda \alpha^{*}_{R}(\theta) k \ ,
\end{aligned}
 \label{eq:5}
\end{equation}

\noindent where $\lambda$ refers to the electron spin in the upper (+) or lower (-) branch, $\theta$ denotes the angle of wave vector with respect to $x$-axis.  $R(\theta)$, the orientation parameter, is defined as \cite{Samira:drag2016}:
   
   \begin{gather}
   R(\theta)=\bigg(\frac{\cos^{2}\theta}{m_{x}}+\frac{\sin^{2}\theta}{m_{y}}\bigg) \ .
   \label{eq:6}
   \end{gather}
 
 \noindent and $ \alpha^{*}_{R} $ is the anisotropic Rashba coefficient, given by: 
  
  \begin{gather}
  \alpha^{*}_{R}(\theta)=\alpha_{R} \sqrt{m_{d} R(\theta)} \ .
  \label{eq:7}
  \end{gather}
  
  \noindent To illustrate the effect of Rashba spin-orbit coupling on the anisotropic band structure, we plot the energy dispersion of the Rashba spin-split branches and energy spacing (the energy difference between the spin up and spin down branches) along the two main crystallographic directions in Fig. \ref{fig1}. Due to the Rashba interaction, the energy dispersion deviates from a parabola for each spin branch. Moreover, the anisotropic characteristic of the spin-split branches as a result of different effective masses along armchair ($x$) and zigzag ($y$) directions is clearly observed. Also, it can be seen, Fig. \ref{fig1}(c) that values of the energy spacing are direction-dependent and as expected along the armchair direction with smaller effective mass, the spin splitting is stronger. It is known that the Rashba spin-orbit interaction can be tuned by an external electric field, so, we show the energy dispersion and energy spacing along the armchair ($x$) direction for different Rashba parameters in Figs. \ref{fig2}(a) and (b). We find that the dispersion relation of a 2D material with anisotropic band structure can be well-tuned by the directional effective masses as well as the Rashba coefficient. Besides, there is a uniform enhancement of the energy spacing with increasing Rashba parameter due to the linear momentum dependence of the spin splitting interaction.
   
\section{Optical Conductivity}
\label{theory2}

 In a spin-orbit coupled system, the optical conductivity due to the transitions between different spin states is an important quantity. We shall calculate this property using the Kubo linear response formalism for a 2D system with anisotropic parabolic energy band in the presence of Rashba interaction. Assuming a spatially homogeneous electric field, the Kubo formula for conductivity which starts from the current-current correlation function is given by \cite{magarill2001spin,rashba1960properties}:

\begin{equation}
\begin{aligned}
\sigma_{ij}(\mathbf{q},t)&=\frac{i n e^{2}}{m \omega} \delta_{ij}
\\&+ \frac{1}{\omega} \int_{0}^{\infty} e^{i \bar{\omega}} <[\hat{J}_{i}(\mathbf{q},t),\hat{J}_{j}(\mathbf{q},0)]>  dt \ ,
\end{aligned}
\label{kubo}
\end{equation}
where indices $i$ and $j$ stand for the two Cartesian coordinates $x, y$, $n$ is the electron density, $\bar{\omega}=\omega+i\eta$ ($\eta \to 0^{+}$) and $\hat{J}_{i}=e \hat{v}_i$ is the current density operator with $\hat{v}_i=\hbar^{-1} \partial \hat{H}/\partial k_{i} $ being the electron velocity operator. In this paper, we concentrate on the absorptive part of the optical conductivity which is the real part of this complex quantity \cite{PhysRevB.87.125425}. In the optical limit $q \to 0 $, the dynamical optical conductivity can be obtained as follows:

\begin{equation}
\begin{aligned}
\sigma_{ij}(\omega)=\frac{i e^{2}}{\omega} &\sum_{\lambda,\lambda^{\prime}}\sum_{\mathbf{k},\mathbf{k^{\prime}}} \psi_{ij}(\mathbf{k},\mathbf{k^{\prime}},\lambda, \lambda^{\prime}) \\
	&\times \frac{f^{0}(E^{\lambda}(\mathbf{k}))-f^{0}(E^{\lambda^{\prime}} (\mathbf{k^{\prime} }))}{E^{\lambda}(\mathbf{k})-E^{\lambda^{\prime}} (\mathbf{k^{\prime}})+\hbar(\omega+i\eta)} \ ,  
\label{eq:8}
\end{aligned}	
\end{equation}

       	  	\begin{figure}[t]
       	  		\includegraphics[width=9.0cm]{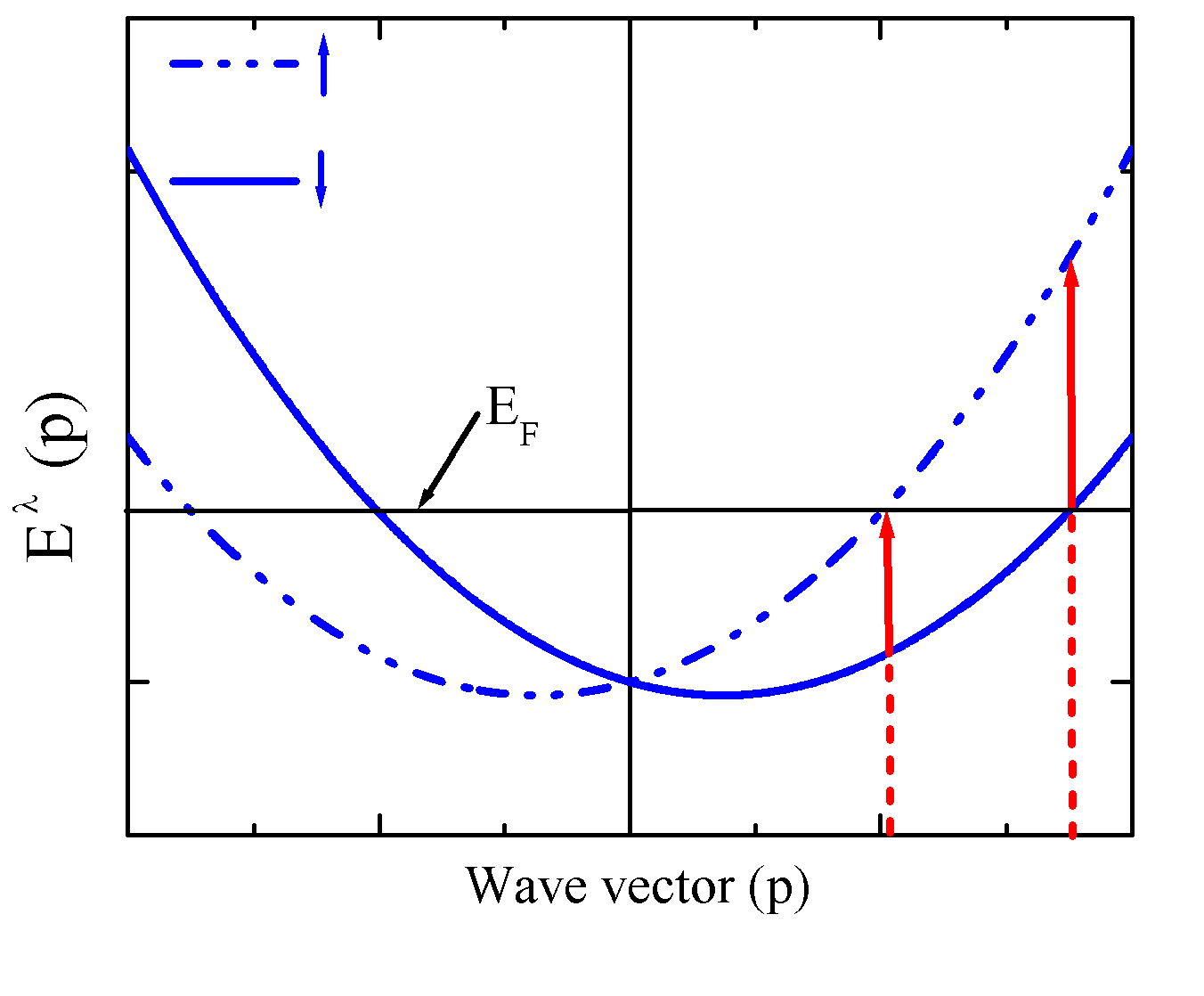}
       	  		\caption{Scheme of transitions between spin branches.
       	  			The arrows correspond to the transitions with threshold frequencies of $(2 \alpha_R p^{\pm}_{F})/\hbar$.}
       	  		\label{fig3}
       	  	\end{figure}

\noindent where $ \psi_{ij}(\mathbf{k},\mathbf{k^{\prime}},\lambda, \lambda^{\prime})$ is defined as:

\begin{equation}
 \psi_{ij}(\mathbf{k},\mathbf{k^{\prime}},\lambda, \lambda^{\prime})=\bra{\mathbf{k}\lambda}\hat{v}_{i} \ket{\mathbf{k^{\prime}} \lambda^{\prime}}  \bra{\mathbf{k^{\prime}} \lambda^{\prime}} \hat{v}_{j} \ket{\mathbf{k} \lambda} \ . 
 \label{eq:9}
 \end{equation}
 
 The electron velocity operators for two spin-split branches in $x$ and $y$ directions are given as:

  \begin{equation}
  \begin{aligned}
 \hat{v}_{x}=\frac{1}{\hbar} \bigg (\frac{\hbar^{2} k_x}{m_x}\hat{\sigma}_0-\alpha_R \sqrt{\frac{m_d}{m_x}} k_x \hat{\sigma}_y \bigg) \ ,
  \\
  \\
   \hat{v}_{y}=\frac{1}{\hbar} \bigg (\frac{\hbar^{2} k_y}{m_y}\hat{\sigma}_0+\alpha_R \sqrt{\frac{m_d}{m_y}} k_y \hat{\sigma}_x \bigg) \ .
  \end{aligned}
  \label{eq:10}
  \end{equation}
 
 Before calculating the eigenstates of the system, $\ket{\mathbf{k} \lambda}$, we introduce a new 2D wave vector $\mathbf{p}=(p_x,p_y)$ which is defined as  $\mathbf{k}= \sqrt{M/m_{d}} \ \mathbf{p} $ with $M$ being the mass tensor whose diagonal elements are $m_{x}$ and $m_{y}$ \cite{Low:prl14}. Now, we can rewrite the free electron and Rashba parts of the total Hamiltonian as follow:
	
\begin{equation}
\begin{aligned}
\hat{H_{0}}=\frac{\hbar^2 p^{2}}{2 m_d}\hat{\sigma}_0 \ ,
\end{aligned}
\label{eq:11}
\end{equation}
\begin{equation}
\begin{aligned}
\hat{H}_{R}=-\alpha_{R} (\mathbf{p} \times \hat{z}).\hat{\sigma} \ ,    
\label{eq:12}
\end{aligned}
\end{equation}
where $\hat{\sigma}=\hat{\sigma}_x\hat{i}+\hat{\sigma}_y\hat{j}$. Thus, the two spin-split eigenstates can be identified as:
\begin{gather}
\ket{\mathbf{p} \lambda}= \frac{e^{(i\mathbf{p}.\mathbf{r})}}{\sqrt{2}} \left
(\begin{array}{cc}
1  \\
\lambda \frac{p_y - ip_x}{p}
\end{array}\right)
\label{eq:13}
\end{gather}   
      	          
\noindent with $r=(x,y)$ being a 2D real space position vector. Also, the expressions for the velocity operators can be written in terms of $\mathbf{p}$:
\\
  \begin{equation}
  \begin{aligned}
 \hat{v}_{x}=\frac{\hbar}{\sqrt{m_x m_d}} \bigg (p_x\hat{\sigma}_0-p_R \hat{\sigma}_y \bigg) \ ,
  \\
  \\
  \hat{v}_{y}=\frac{\hbar}{\sqrt{m_y m_d}} \bigg (p_y\hat{\sigma}_0+p_R \hat{\sigma}_x \bigg) \ ,
  \end{aligned}
  \label{eq:14}
  \end{equation}
where $p_R= \alpha_R m_d /\hbar^2$ is the effective Rashba wave vector.
 It has been shown that the spin-conserving intra-band transitions give rise to low frequency absorption, whereas the spin-flip transitions result in a wide absorption peak\cite{PhysRevB.94.161404,PhysRevB.94.155432}. We focus on the optical conductivity due to the inter-band spin flip excitations. As ordinary 2D electron gas systems \cite{PhysRevB.94.161404}, it can be shown that the off-diagonal elements of the optical conductivity (transverse or Hall conductivity) are zero in the absence of a magnetic field, \textit{i.e.} $\sigma_{xy}=0$. By making use of Eqs. (\ref{eq:13}) and (\ref{eq:14}), one can calculate the diagonal elements of $\psi$ tensor as:
      	  	     
 \begin{widetext}
  \begin{equation}
  \begin{aligned}
   \psi _{jj}(\mathbf{p}, \mathbf{p^{\prime}};\lambda, \lambda^{\prime})=\frac{\hbar^2}{2m_{j}m_{d}} \bigg (p_j^2(1+\lambda \lambda^{\prime})+ p^{2}_{\alpha}(1+\lambda \lambda^{\prime} cos 2\phi)+2p_R p_j cos\phi (1+\lambda \lambda^{\prime}) \bigg) \delta_{\mathbf{p} \mathbf{p^{\prime}}} \ ,
   \end{aligned}
  \label{eq:15}
  \end{equation} 
\end{widetext}          	 
  where $\phi=tan^{-1}(p_{y}/p_{x})$. Accordingly, the non-equal diagonal components of optical conductivity (longitudinal optical conductivity) due to the spin-flip transitions are obtained as:
  
  \begin{widetext}
  	\begin{equation}
  	\begin{aligned}
  	& \sigma _{jj}(\omega)=\frac{i e^{2}\hbar^2 p_R}{2 \omega m_{j}m_{d}} \int \frac{pdp d\phi}{(2\pi)^{2}} (1-cos 2\phi)  \bigg(\frac{f^{0}(E^{-}(\mathbf{p}))-f^{0}(E^{+}(\mathbf{p }))}{E^{-}(\mathbf{p})-E^{+}(\mathbf{p})+\hbar(\omega+i\eta)}+ (E^{-} \leftrightarrow E^{+}) \bigg) \ .
 	\end{aligned}
  	\label{eq:16}
  	\end{equation} 
  \end{widetext}  
                         
      \begin{figure}[hb]
      	\includegraphics[width=8.50cm]{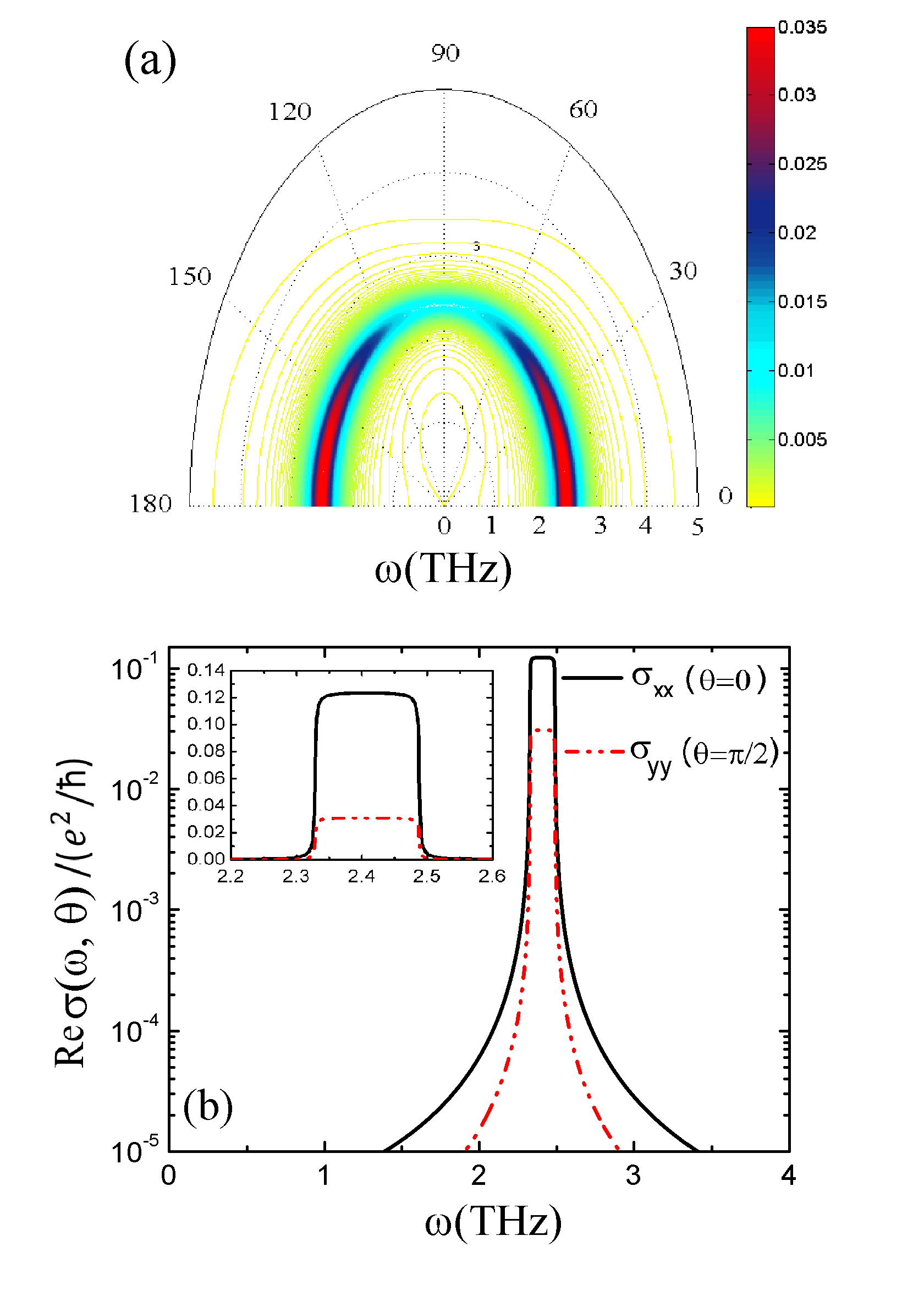}
      	\caption{Real part of longitudinal optical conductivity as a function of excitation frequency: (a) for all $\theta$ (b) for $\theta=0$ and $\theta=\pi/2$ with $n = 1\times10^{13}$cm$^{-2}$. Here we set $\alpha_{R}$=10 meV$\mathring{A}$, $m_y$=4$m_0$ and $m_x$=$m_0$.}
      	\label{fig4}
      \end{figure} 
      
 \noindent After performing the $\mathbf{p}$ integral, the following expression is obtained for $\sigma _{jj}$ at zero temperature:

 	\begin{equation}
  \sigma _{jj}(\omega)= -\frac{i e^{2} m^2_d}{2 \pi m_j \omega \hbar^4} \bigg[ 1+ \frac{\hbar^3 \omega R(\theta) \Lambda (\omega)}{8 
  	(\alpha^{*}_{R}(\theta))^2 }\bigg] \ ,
 	\label{eq:17}
 	\end{equation}
 	
 \noindent with $\Lambda (\omega) $ defined as:

 	\begin{equation}
 	\Lambda(\omega)=\ln\bigg(\frac{(\omega-\omega_{-}+i\eta)(\omega+\omega_{+}+i\eta)}{(\omega+\omega_{-}+i\eta)(\omega-\omega_{+}+i\eta)}\bigg) \ ,
 	\label{eq:18}
 	\end{equation}

\noindent where $\omega_{\pm}$ are the threshold frequency modes induced by inter-branch electronic transition:  
 
\begin{gather}
\omega_{\pm}=\frac{2 \alpha^{*}_{R} p^{F}_{\pm}}{\sqrt{m_d R(\theta)}\hbar} \ .  %\ m_{d} R(\theta) %
\label{eq:19}
\end{gather}

 \noindent $p^{F}_{\pm}= p_{F} \mp p_R$ are the Fermi wave vectors for the two spin branches with $p_{F}=\sqrt{2n\pi-p^{2}_R}$. A schematic diagram for the optical transitions are shown in Fig. \ref{fig3}. The arrows correspond to the vertical transitions between two spin branches in the optical limit ($q \to 0$).

 The longitudinal optical conductivity along an arbitrary polarization direction $\theta$, is given by \cite{PhysRevB.90.075434}:

\begin{equation}
\sigma(\omega,\theta)= \sigma_{xx}  cos^2\theta+ \sigma_{yy}  sin^2\theta \ .
\label{eq:20}
\end{equation}

\noindent By inserting Eqs. (\ref{eq:17}) and (\ref{eq:18}) in the above equation we obtain:

\begin{equation}
\sigma(\omega,\theta)=-\frac{i e^{2} m_d (\alpha^{*}_{R}(\theta))^2} {2 \pi \omega \hbar^4} \bigg[ 1+ \frac{\hbar^3 \omega R(\theta) \Lambda (\omega)}{8 
	(\alpha^{*}_{R}(\theta))^2 }\bigg] \ .
\label{eq:21}
\end{equation}

\begin{figure*}
	\includegraphics[width=18.cm]{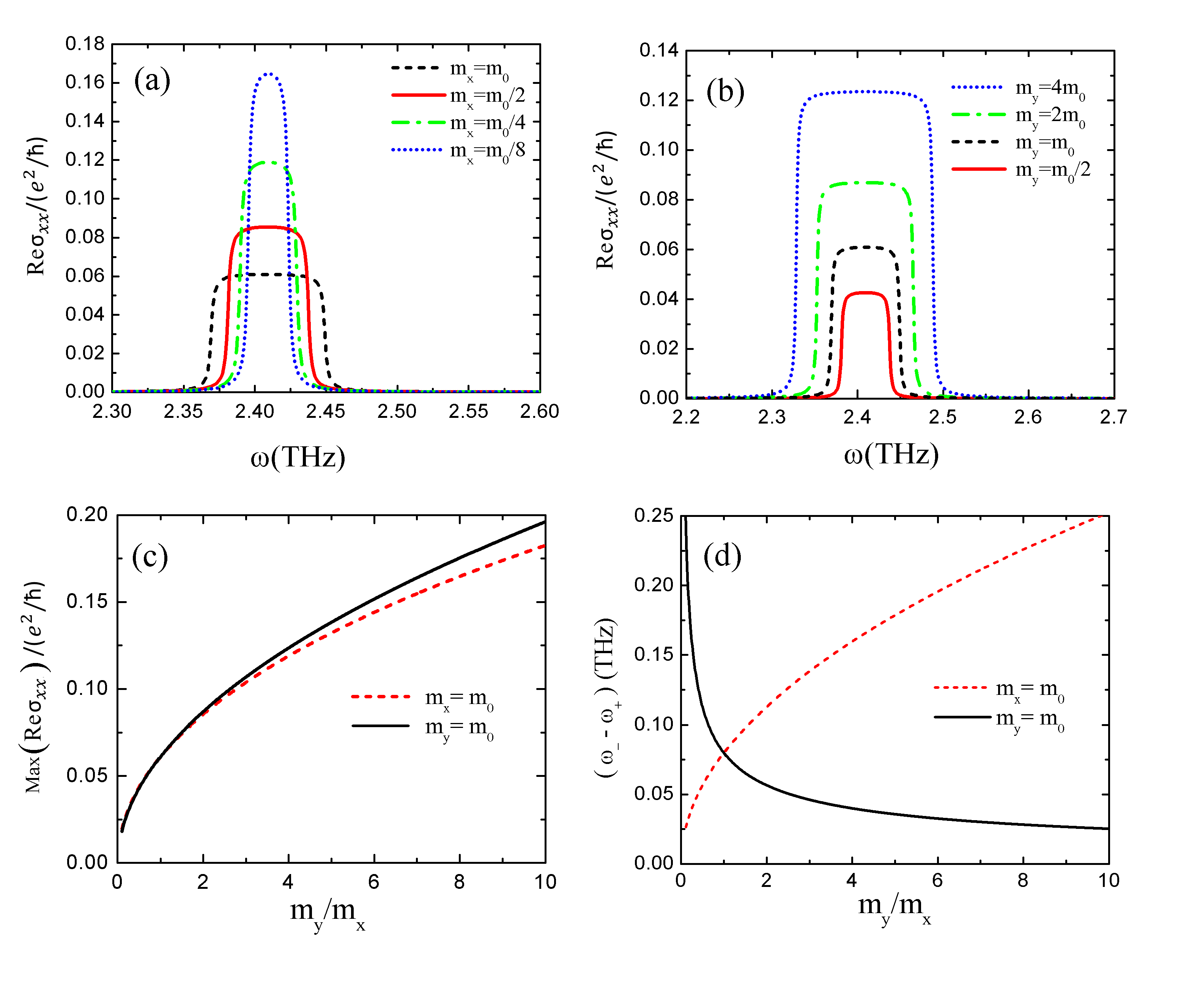}
	\caption{Real part of longitudinal optical conductivity  as a function of excitation frequency for $\theta=0$ with (a) $m_y=4m_0$ and (b) $m_x=m_0$. (c) Maximum of real part of longitudinal optical conductivity as a function of mass anisotropy ratio. (d) Width of the absorption window as a function of mass anisotropy ratio. Here, we set $n = 1\times10^{13}$cm$^{-2}$ with $\alpha_{R}$=10 meV $\mathring{A}$ .}
	\label{fig5}
\end{figure*}

In Fig. \ref{fig4}(a), we illustrate the calculated real part of the anisotropic conductivity for arbitrary direction of polarization vector as a function of radiation frequency, $\omega$, at a fixed electron density $n = 1\times10^{13}$cm$^{-2}$ and for $\alpha_{R}$=10 meV$\mathring{A}$, $m_y$=4$m_0$ and $m_x$=$m_0$. As expected, the optical conductivity has its maximum value at $\theta=0$ \textit{i.e.} along the armchair direction of the 2D layer. The significantly smaller effective mass for $\theta=0$ direction suggests that the charge carriers prefer to flow along this direction. In addition, Fig. \ref{fig4}(b) shows the absorption part of the optical conductivity for two main crystallographic directions $\theta=0$ and $\theta=\pi/2$. The fact that the anisotropy ratio of the optical conductivity is equal to the inverse of the mass anisotropy ratio, \textit{i.e.} $\sigma _{yy}(\omega)/\sigma _{xx}(\omega)= m_x/ m_y $, is clearly observed in this figure.	 {\color{black}Furthermore, as a consequence of the energy conservation law under vertical transitions ($q \to 0$) between spin branches (see Fig. \ref{fig3})),  the absorption part of the longitudinal optical conductivity at $T = 0$ has a step function like variation with the frequency, \textit{i.e.} a non-zero value for a range of frequency which is given by:

\begin{equation}
\omega_{-} -\omega_{+}= \frac{2 (\alpha^{*}_{R})^2 }{\hbar^{3} R(\theta)} \ .  %\ m_{d} R(\theta) %
\label{eq:22}
\end{equation} 
  As is evident from Fig. \ref{fig4}, the absorption is in the THz region for the set of parameters used here.
\begin{figure*}[ht]
	\includegraphics[width=18.0cm]{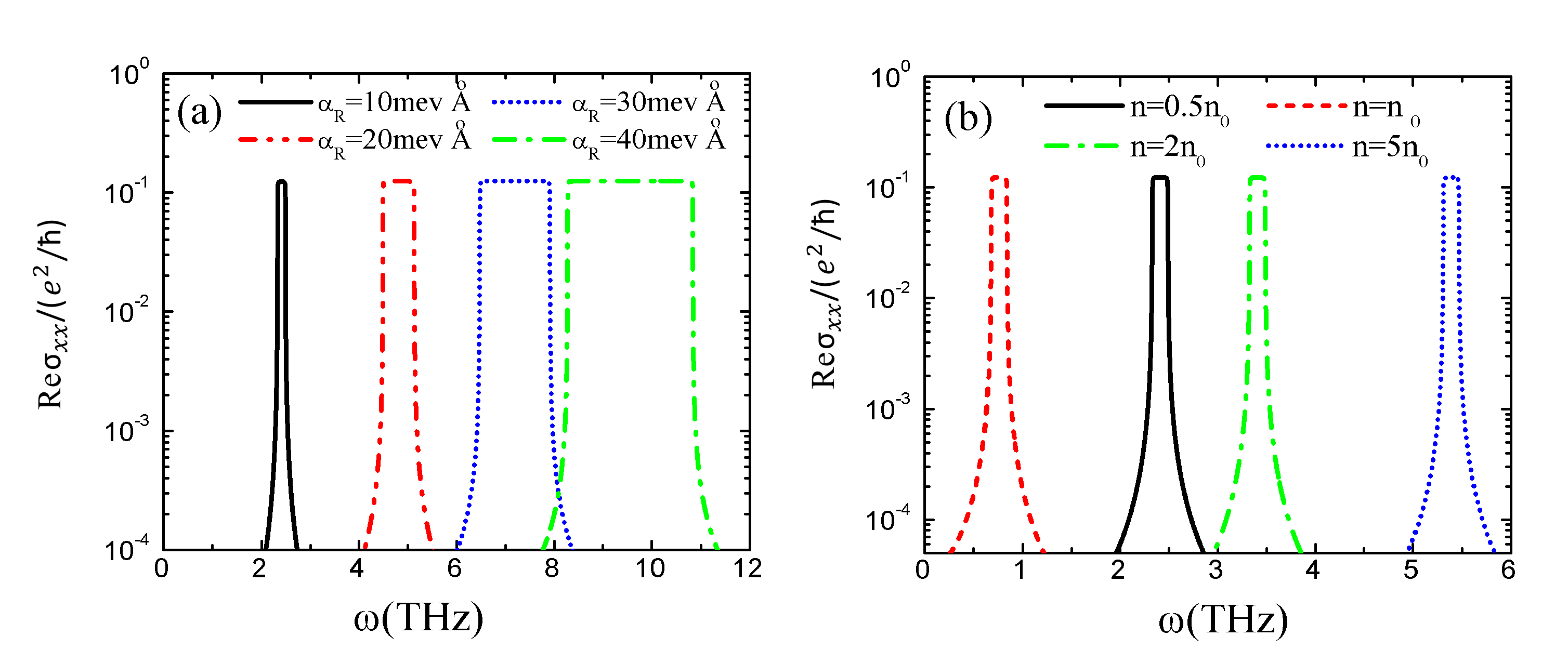}
	\caption{Real part of longitudinal optical conductivity as a function of excitation frequency for $\theta$=0 with $m_y$=4$m_0$, $m_x$=$m_0$ for a few values of (a) Rashba parameter with  $n_{0} = 1\times10^{13}$cm$^{-2}$ (b) electron density with $\alpha_{R}$=10 meV  $\mathring{A}$ }. 
	\label{fig6}
\end{figure*}  
We have already mentioned above that the optical conductivity exhibits a strong dependence on the effective mass anisotropy. The variations of the real part of $\sigma_{xx} $ with the change of effective mass along the $x$ and $y$ directions are shown in Fig. \ref{fig5}. One may notice that the mass asymmetry not only alters the maximum value of the optical conductivity but also changes the width of the peak.  According to this figure, the peak height of $\sigma _{xx}(\omega)$ increases by decreasing the effective mass along the radiation polarization direction or increasing the effective mass along the direction perpendicular to the radiation polarization. In other words, a higher peak intensity is achieved as a consequence of the effective mass anisotropy ratio ($ m_y/ m_x $) enhancement.  Moreover, the absorption window of the real part of longitudinal optical conductivity is extended by increasing the density of states mass ($m_d$).

The effect of Rashba coupling strength on the absorption part of $\sigma _{xx}(\omega)$ is shown in Fig. \ref{fig6}(a). The absorption peak width is broadened and the onset frequency of the absorptive peak moves towards higher value when the Rashba parameter increases \cite{PhysRevB.72.033320}. The Rashba parameters chosen here are comparable to the as-obtained values for phosphorene and group-IV monochalcogenides from density functional theory calculations \cite{PhysRevB.92.035135,PhysRevB.94.155423}. For large values of $\alpha_{R}$, a stronger spin splitting occurs which in turn shifts the absorption part of the optical conductivity from the THz to infrared frequencies. These features for the spin-flip absorption peak are similar to that  of the conventional two dimensional electron gas system and \MS \cite{PhysRevB.94.155432}. We also depict the variation of  the optical absorption with the electron density in Fig. \ref{fig6}(b) for fixed $\alpha_{R}$=10 meV  $\mathring{A}$ and effective masses $m_y$=4$m_0$ and $m_x$=$m_0$ . One of the important characteristics of this figure is that while the value of the peak height remains almost unchanged by increasing electron density, the absorption peak moves toward higher frequencies as a result of the Pauli blockade effect\cite{PhysRevLett.111.246601,1751-8121-42-21-214033,PhysRevLett.97.076403}.

\section{conclusion} \label{conclusion}

In summary, we have studied the energy spectrum and optical response of an anisotropic 2D electron gas system in the presence of Rashba spin-orbit interaction. Based on the Kubo formalism, we calculated the optical conductivity tensor considering the Rashba spin-flip excitations. We found that the effective mass anisotropy plays an important role in the optical absorption spectrum through the direction-dependent Rashba spin splitting. As a general result, the diagonal components of the optical conductivity tensor are inversely proportional to the corresponding effective mass elements. Furthermore, the effective mass asymmetry is an additional degree of freedom to tune the height and width of the absorption peak. This introduces new aspects to the optical conductivity for spintronic applications of 2D anisotropic materials such as phosphorene and group-IV monochalcogenides.  We also showed that larger optical absorption is generated when the polarization of radiation is along the armchair direction and its maximal value is enhanced by increasing the effective mass ratio. However, the width of the absorption window has a strong dependence on both the polarization direction and the effective mass ratio.  Finally, the position of absorptive peak moves to higher frequencies with increasing the Rashba parameter and electron density.  Our results suggest an interesting way to determine some of the spintronic characteristics of a class of 2D nanostructures, with anisotropic Rashba effect, using optical methods.  
 
%\bibliography{Ref}

\begin{thebibliography}{48}%
		\makeatletter
		\providecommand \@ifxundefined [1]{%
			\@ifx{#1\undefined}
		}%
		\providecommand \@ifnum [1]{%
			\ifnum #1\expandafter \@firstoftwo
			\else \expandafter \@secondoftwo
			\fi
		}%
		\providecommand \@ifx [1]{%
			\ifx #1\expandafter \@firstoftwo
			\else \expandafter \@secondoftwo
			\fi
		}%
		\providecommand \natexlab [1]{#1}%
		\providecommand \enquote  [1]{``#1''}%
		\providecommand \bibnamefont  [1]{#1}%
		\providecommand \bibfnamefont [1]{#1}%
		\providecommand \citenamefont [1]{#1}%
		\providecommand \href@noop [0]{\@secondoftwo}%
		\providecommand \href [0]{\begingroup \@sanitize@url \@href}%
		\providecommand \@href[1]{\@@startlink{#1}\@@href}%
		\providecommand \@@href[1]{\endgroup#1\@@endlink}%
		\providecommand \@sanitize@url [0]{\catcode `\\12\catcode `\$12\catcode
			`\&12\catcode `\#12\catcode `\^12\catcode `\_12\catcode `\%12\relax}%
		\providecommand \@@startlink[1]{}%
		\providecommand \@@endlink[0]{}%
		\providecommand \url  [0]{\begingroup\@sanitize@url \@url }%
		\providecommand \@url [1]{\endgroup\@href {#1}{\urlprefix }}%
		\providecommand \urlprefix  [0]{URL }%
		\providecommand \Eprint [0]{\href }%
		\providecommand \doibase [0]{http://dx.doi.org/}%
		\providecommand \selectlanguage [0]{\@gobble}%
		\providecommand \bibinfo  [0]{\@secondoftwo}%
		\providecommand \bibfield  [0]{\@secondoftwo}%
		\providecommand \translation [1]{[#1]}%
		\providecommand \BibitemOpen [0]{}%
		\providecommand \bibitemStop [0]{}%
		\providecommand \bibitemNoStop [0]{.\EOS\space}%
		\providecommand \EOS [0]{\spacefactor3000\relax}%
		\providecommand \BibitemShut  [1]{\csname bibitem#1\endcsname}%
		\let\auto@bib@innerbib\@empty
		%</preamble>
		\bibitem [{\citenamefont {Carvalho}\ \emph {et~al.}(2016)\citenamefont
			{Carvalho}, \citenamefont {Wang}, \citenamefont {Zhu}, \citenamefont {Rodin},
			\citenamefont {Su},\ and\ \citenamefont {Neto}}]{Phosphorene:review}%
		\BibitemOpen
		\bibfield  {author} {\bibinfo {author} {\bibfnamefont {A.}~\bibnamefont
				{Carvalho}}, \bibinfo {author} {\bibfnamefont {M.}~\bibnamefont {Wang}},
			\bibinfo {author} {\bibfnamefont {X.}~\bibnamefont {Zhu}}, \bibinfo {author}
			{\bibfnamefont {A.~S.}\ \bibnamefont {Rodin}}, \bibinfo {author}
			{\bibfnamefont {H.}~\bibnamefont {Su}}, \ and\ \bibinfo {author}
			{\bibfnamefont {A.~H.~C.}\ \bibnamefont {Neto}},\ }\href {\doibase
			10.1038/natrevmats.2016.61} {\bibfield  {journal} {\bibinfo  {journal}
				{Nature Reviews Materials}\ }\textbf {\bibinfo {volume} {1}} (\bibinfo {year}
			{2016}),\ 10.1038/natrevmats.2016.61}\BibitemShut {NoStop}%
		\bibitem [{\citenamefont {Ling}\ \emph {et~al.}(2015)\citenamefont {Ling},
			\citenamefont {Wang}, \citenamefont {Huang}, \citenamefont {Xia},\ and\
			\citenamefont {Dresselhaus}}]{Ling:pnas15}%
		\BibitemOpen
		\bibfield  {author} {\bibinfo {author} {\bibfnamefont {X.}~\bibnamefont
				{Ling}}, \bibinfo {author} {\bibfnamefont {H.}~\bibnamefont {Wang}}, \bibinfo
			{author} {\bibfnamefont {S.}~\bibnamefont {Huang}}, \bibinfo {author}
			{\bibfnamefont {F.}~\bibnamefont {Xia}}, \ and\ \bibinfo {author}
			{\bibfnamefont {M.~S.}\ \bibnamefont {Dresselhaus}},\ }\href {\doibase
			10.1073/pnas.1416581112} {\bibfield  {journal} {\bibinfo  {journal} {Proc.
					Natl. Acad. Sci. U.S.A.}\ }\textbf {\bibinfo {volume} {112}},\ \bibinfo
			{pages} {4523} (\bibinfo {year} {2015})}\BibitemShut {NoStop}%
		\bibitem [{\citenamefont {Taghizadeh~Sisakht}\ \emph
			{et~al.}(2016)\citenamefont {Taghizadeh~Sisakht}, \citenamefont {Fazileh},
			\citenamefont {Zare}, \citenamefont {Zarenia},\ and\ \citenamefont
			{Peeters}}]{PhysRevB.94.085417}%
		\BibitemOpen
		\bibfield  {author} {\bibinfo {author} {\bibfnamefont {E.}~\bibnamefont
				{Taghizadeh~Sisakht}}, \bibinfo {author} {\bibfnamefont {F.}~\bibnamefont
				{Fazileh}}, \bibinfo {author} {\bibfnamefont {M.~H.}\ \bibnamefont {Zare}},
			\bibinfo {author} {\bibfnamefont {M.}~\bibnamefont {Zarenia}}, \ and\
			\bibinfo {author} {\bibfnamefont {F.~M.}\ \bibnamefont {Peeters}},\ }\href
		{\doibase 10.1103/PhysRevB.94.085417} {\bibfield  {journal} {\bibinfo
				{journal} {Phys. Rev. B}\ }\textbf {\bibinfo {volume} {94}},\ \bibinfo
			{pages} {085417} (\bibinfo {year} {2016})}\BibitemShut {NoStop}%
		\bibitem [{\citenamefont {Zare}\ \emph {et~al.}(2017)\citenamefont {Zare},
			\citenamefont {Rameshti}, \citenamefont {Ghamsari},\ and\ \citenamefont
			{Asgari}}]{PhysRevB.95.045422}%
		\BibitemOpen
		\bibfield  {author} {\bibinfo {author} {\bibfnamefont {M.}~\bibnamefont
				{Zare}}, \bibinfo {author} {\bibfnamefont {B.~Z.}\ \bibnamefont {Rameshti}},
			\bibinfo {author} {\bibfnamefont {F.~G.}\ \bibnamefont {Ghamsari}}, \ and\
			\bibinfo {author} {\bibfnamefont {R.}~\bibnamefont {Asgari}},\ }\href
		{\doibase 10.1103/PhysRevB.95.045422} {\bibfield  {journal} {\bibinfo
				{journal} {Phys. Rev. B}\ }\textbf {\bibinfo {volume} {95}},\ \bibinfo
			{pages} {045422} (\bibinfo {year} {2017})}\BibitemShut {NoStop}%
		\bibitem [{\citenamefont {Brener}\ \emph {et~al.}(2017)\citenamefont {Brener},
			\citenamefont {Rudenko},\ and\ \citenamefont
			{Katsnelson}}]{PhysRevB.95.041406}%
		\BibitemOpen
		\bibfield  {author} {\bibinfo {author} {\bibfnamefont {S.}~\bibnamefont
				{Brener}}, \bibinfo {author} {\bibfnamefont {A.~N.}\ \bibnamefont {Rudenko}},
			\ and\ \bibinfo {author} {\bibfnamefont {M.~I.}\ \bibnamefont {Katsnelson}},\
		}\href {\doibase 10.1103/PhysRevB.95.041406} {\bibfield  {journal} {\bibinfo
			{journal} {Phys. Rev. B}\ }\textbf {\bibinfo {volume} {95}},\ \bibinfo
		{pages} {041406} (\bibinfo {year} {2017})}\BibitemShut {NoStop}%
	\bibitem [{\citenamefont {Tran}\ \emph {et~al.}(2014)\citenamefont {Tran},
		\citenamefont {Soklaski}, \citenamefont {Liang},\ and\ \citenamefont
		{Yang}}]{PhysRevB.89.235319}%
	\BibitemOpen
	\bibfield  {author} {\bibinfo {author} {\bibfnamefont {V.}~\bibnamefont
			{Tran}}, \bibinfo {author} {\bibfnamefont {R.}~\bibnamefont {Soklaski}},
		\bibinfo {author} {\bibfnamefont {Y.}~\bibnamefont {Liang}}, \ and\ \bibinfo
		{author} {\bibfnamefont {L.}~\bibnamefont {Yang}},\ }\href {\doibase
		10.1103/PhysRevB.89.235319} {\bibfield  {journal} {\bibinfo  {journal} {Phys.
				Rev. B}\ }\textbf {\bibinfo {volume} {89}},\ \bibinfo {pages} {235319}
		(\bibinfo {year} {2014})}\BibitemShut {NoStop}%
	\bibitem [{\citenamefont {Sa}\ \emph {et~al.}(2015)\citenamefont {Sa},
		\citenamefont {Li}, \citenamefont {Sun}, \citenamefont {Qi}, \citenamefont
		{Wen},\ and\ \citenamefont {Wu}}]{0957-4484-26-21-215205}%
	\BibitemOpen
	\bibfield  {author} {\bibinfo {author} {\bibfnamefont {B.}~\bibnamefont
			{Sa}}, \bibinfo {author} {\bibfnamefont {Y.-L.}\ \bibnamefont {Li}}, \bibinfo
		{author} {\bibfnamefont {Z.}~\bibnamefont {Sun}}, \bibinfo {author}
		{\bibfnamefont {J.}~\bibnamefont {Qi}}, \bibinfo {author} {\bibfnamefont
			{C.}~\bibnamefont {Wen}}, \ and\ \bibinfo {author} {\bibfnamefont
			{B.}~\bibnamefont {Wu}},\ }\href
	{http://stacks.iop.org/0957-4484/26/i=21/a=215205} {\bibfield  {journal}
		{\bibinfo  {journal} {Nanotechnology}\ }\textbf {\bibinfo {volume} {26}},\
		\bibinfo {pages} {215205} (\bibinfo {year} {2015})}\BibitemShut {NoStop}%
	\bibitem [{\citenamefont {Zhang}\ \emph {et~al.}(2016)\citenamefont {Zhang},
		\citenamefont {Wang}, \citenamefont {Liu}, \citenamefont {Huang},
		\citenamefont {Zhou}, \citenamefont {Cai}, \citenamefont {Xie}, \citenamefont
		{Yang}, \citenamefont {Chen},\ and\ \citenamefont
		{Zeng}}]{0957-4484-27-27-274001}%
	\BibitemOpen
	\bibfield  {author} {\bibinfo {author} {\bibfnamefont {S.}~\bibnamefont
			{Zhang}}, \bibinfo {author} {\bibfnamefont {N.}~\bibnamefont {Wang}},
		\bibinfo {author} {\bibfnamefont {S.}~\bibnamefont {Liu}}, \bibinfo {author}
		{\bibfnamefont {S.}~\bibnamefont {Huang}}, \bibinfo {author} {\bibfnamefont
			{W.}~\bibnamefont {Zhou}}, \bibinfo {author} {\bibfnamefont {B.}~\bibnamefont
			{Cai}}, \bibinfo {author} {\bibfnamefont {M.}~\bibnamefont {Xie}}, \bibinfo
		{author} {\bibfnamefont {Q.}~\bibnamefont {Yang}}, \bibinfo {author}
		{\bibfnamefont {X.}~\bibnamefont {Chen}}, \ and\ \bibinfo {author}
		{\bibfnamefont {H.}~\bibnamefont {Zeng}},\ }\href
	{http://stacks.iop.org/0957-4484/27/i=27/a=274001} {\bibfield  {journal}
		{\bibinfo  {journal} {Nanotechnology}\ }\textbf {\bibinfo {volume} {27}},\
		\bibinfo {pages} {274001} (\bibinfo {year} {2016})}\BibitemShut {NoStop}%
	\bibitem [{\citenamefont {Gomes}\ and\ \citenamefont
		{Carvalho}(2015)}]{PhysRevB.92.085406}%
	\BibitemOpen
	\bibfield  {author} {\bibinfo {author} {\bibfnamefont {L.~C.}\ \bibnamefont
			{Gomes}}\ and\ \bibinfo {author} {\bibfnamefont {A.}~\bibnamefont
			{Carvalho}},\ }\href {\doibase 10.1103/PhysRevB.92.085406} {\bibfield
		{journal} {\bibinfo  {journal} {Phys. Rev. B}\ }\textbf {\bibinfo {volume}
			{92}},\ \bibinfo {pages} {085406} (\bibinfo {year} {2015})}\BibitemShut
	{NoStop}%
	\bibitem [{\citenamefont {Kamal}\ \emph {et~al.}(2016)\citenamefont {Kamal},
		\citenamefont {Chakrabarti},\ and\ \citenamefont
		{Ezawa}}]{PhysRevB.93.125428}%
	\BibitemOpen
	\bibfield  {author} {\bibinfo {author} {\bibfnamefont {C.}~\bibnamefont
			{Kamal}}, \bibinfo {author} {\bibfnamefont {A.}~\bibnamefont {Chakrabarti}},
		\ and\ \bibinfo {author} {\bibfnamefont {M.}~\bibnamefont {Ezawa}},\ }\href
	{\doibase 10.1103/PhysRevB.93.125428} {\bibfield  {journal} {\bibinfo
			{journal} {Phys. Rev. B}\ }\textbf {\bibinfo {volume} {93}},\ \bibinfo
		{pages} {125428} (\bibinfo {year} {2016})}\BibitemShut {NoStop}%
	\bibitem [{\citenamefont {Medrano~Sandonas}\ \emph {et~al.}(2016)\citenamefont
		{Medrano~Sandonas}, \citenamefont {Teich}, \citenamefont {Gutierrez},
		\citenamefont {Lorenz}, \citenamefont {Pecchia}, \citenamefont {Seifert},\
		and\ \citenamefont {Cuniberti}}]{Med:JPC16}%
	\BibitemOpen
	\bibfield  {author} {\bibinfo {author} {\bibfnamefont {L.}~\bibnamefont
			{Medrano~Sandonas}}, \bibinfo {author} {\bibfnamefont {D.}~\bibnamefont
			{Teich}}, \bibinfo {author} {\bibfnamefont {R.}~\bibnamefont {Gutierrez}},
		\bibinfo {author} {\bibfnamefont {T.}~\bibnamefont {Lorenz}}, \bibinfo
		{author} {\bibfnamefont {A.}~\bibnamefont {Pecchia}}, \bibinfo {author}
		{\bibfnamefont {G.}~\bibnamefont {Seifert}}, \ and\ \bibinfo {author}
		{\bibfnamefont {G.}~\bibnamefont {Cuniberti}},\ }\href
	{http://pubs.acs.org/doi/abs/10.1021/acs.jpcc.6b04969} {\bibfield  {journal}
		{\bibinfo  {journal} {The Journal of Physical Chemistry C}\ }\textbf
		{\bibinfo {volume} {120}},\ \bibinfo {pages} {18841} (\bibinfo {year}
		{2016})}\BibitemShut {NoStop}%
	\bibitem [{\citenamefont {Tian}\ \emph {et~al.}(2017)\citenamefont {Tian},
		\citenamefont {Guo}, \citenamefont {Zhao}, \citenamefont {Li},\ and\
		\citenamefont {Xue}}]{2DSnS}%
	\BibitemOpen
	\bibfield  {author} {\bibinfo {author} {\bibfnamefont {Z.}~\bibnamefont
			{Tian}}, \bibinfo {author} {\bibfnamefont {C.}~\bibnamefont {Guo}}, \bibinfo
		{author} {\bibfnamefont {M.}~\bibnamefont {Zhao}}, \bibinfo {author}
		{\bibfnamefont {R.}~\bibnamefont {Li}}, \ and\ \bibinfo {author}
		{\bibfnamefont {J.}~\bibnamefont {Xue}},\ }\href {\doibase
		10.1021/acsnano.6b08704} {\bibfield  {journal} {\bibinfo  {journal} {ACS
				Nano}\ }\textbf {\bibinfo {volume} {11}},\ \bibinfo {pages} {2219} (\bibinfo
		{year} {2017})}\BibitemShut {NoStop}%
	\bibitem [{\citenamefont {\ifmmode \check{Z}\else
			\v{Z}\fi{}uti\ifmmode~\acute{c}\else \'{c}\fi{}}\ \emph
		{et~al.}(2004)\citenamefont {\ifmmode \check{Z}\else
			\v{Z}\fi{}uti\ifmmode~\acute{c}\else \'{c}\fi{}}, \citenamefont {Fabian},\
		and\ \citenamefont {Das~Sarma}}]{RevModPhys.76.323}%
	\BibitemOpen
	\bibfield  {author} {\bibinfo {author} {\bibfnamefont {I.}~\bibnamefont
			{\ifmmode \check{Z}\else \v{Z}\fi{}uti\ifmmode~\acute{c}\else \'{c}\fi{}}},
		\bibinfo {author} {\bibfnamefont {J.}~\bibnamefont {Fabian}}, \ and\ \bibinfo
		{author} {\bibfnamefont {S.}~\bibnamefont {Das~Sarma}},\ }\href {\doibase
		10.1103/RevModPhys.76.323} {\bibfield  {journal} {\bibinfo  {journal} {Rev.
				Mod. Phys.}\ }\textbf {\bibinfo {volume} {76}},\ \bibinfo {pages} {323}
		(\bibinfo {year} {2004})}\BibitemShut {NoStop}%
	\bibitem [{\citenamefont {Min}\ \emph {et~al.}(2006)\citenamefont {Min},
		\citenamefont {Hill}, \citenamefont {Sinitsyn}, \citenamefont {Sahu},
		\citenamefont {Kleinman},\ and\ \citenamefont
		{MacDonald}}]{PhysRevB.74.165310}%
	\BibitemOpen
	\bibfield  {author} {\bibinfo {author} {\bibfnamefont {H.}~\bibnamefont
			{Min}}, \bibinfo {author} {\bibfnamefont {J.~E.}\ \bibnamefont {Hill}},
		\bibinfo {author} {\bibfnamefont {N.~A.}\ \bibnamefont {Sinitsyn}}, \bibinfo
		{author} {\bibfnamefont {B.~R.}\ \bibnamefont {Sahu}}, \bibinfo {author}
		{\bibfnamefont {L.}~\bibnamefont {Kleinman}}, \ and\ \bibinfo {author}
		{\bibfnamefont {A.~H.}\ \bibnamefont {MacDonald}},\ }\href {\doibase
		10.1103/PhysRevB.74.165310} {\bibfield  {journal} {\bibinfo  {journal} {Phys.
				Rev. B}\ }\textbf {\bibinfo {volume} {74}},\ \bibinfo {pages} {165310}
		(\bibinfo {year} {2006})}\BibitemShut {NoStop}%
	\bibitem [{\citenamefont {Pesin}\ and\ \citenamefont
		{MacDonald}(2012)}]{SpintronicsNat:MacDonald}%
	\BibitemOpen
	\bibfield  {author} {\bibinfo {author} {\bibfnamefont {D.}~\bibnamefont
			{Pesin}}\ and\ \bibinfo {author} {\bibfnamefont {A.~H.}\ \bibnamefont
			{MacDonald}},\ }\href {\doibase 10.1038/nmat3305} {\bibfield  {journal}
		{\bibinfo  {journal} {Nature Materials}\ }\textbf {\bibinfo {volume} {11}},\
		\bibinfo {pages} {409} (\bibinfo {year} {2012})}\BibitemShut {NoStop}%
	\bibitem [{\citenamefont {Wunderlich}\ \emph {et~al.}(2005)\citenamefont
		{Wunderlich}, \citenamefont {Kaestner}, \citenamefont {Sinova},\ and\
		\citenamefont {Jungwirth}}]{PhysRevLett.94.047204}%
	\BibitemOpen
	\bibfield  {author} {\bibinfo {author} {\bibfnamefont {J.}~\bibnamefont
			{Wunderlich}}, \bibinfo {author} {\bibfnamefont {B.}~\bibnamefont
			{Kaestner}}, \bibinfo {author} {\bibfnamefont {J.}~\bibnamefont {Sinova}}, \
		and\ \bibinfo {author} {\bibfnamefont {T.}~\bibnamefont {Jungwirth}},\ }\href
	{\doibase 10.1103/PhysRevLett.94.047204} {\bibfield  {journal} {\bibinfo
			{journal} {Phys. Rev. Lett.}\ }\textbf {\bibinfo {volume} {94}},\ \bibinfo
		{pages} {047204} (\bibinfo {year} {2005})}\BibitemShut {NoStop}%
	\bibitem [{\citenamefont {Shanavas}\ and\ \citenamefont
		{Satpathy}(2014)}]{PhysRevLett.112.086802}%
	\BibitemOpen
	\bibfield  {author} {\bibinfo {author} {\bibfnamefont {K.~V.}\ \bibnamefont
			{Shanavas}}\ and\ \bibinfo {author} {\bibfnamefont {S.}~\bibnamefont
			{Satpathy}},\ }\href {\doibase 10.1103/PhysRevLett.112.086802} {\bibfield
		{journal} {\bibinfo  {journal} {Phys. Rev. Lett.}\ }\textbf {\bibinfo
			{volume} {112}},\ \bibinfo {pages} {086802} (\bibinfo {year}
		{2014})}\BibitemShut {NoStop}%
	\bibitem [{\citenamefont {Galitski}\ and\ \citenamefont
		{Spielman}(2013)}]{Spinquantumgas:2013}%
	\BibitemOpen
	\bibfield  {author} {\bibinfo {author} {\bibfnamefont {V.}~\bibnamefont
			{Galitski}}\ and\ \bibinfo {author} {\bibfnamefont {I.~B.}\ \bibnamefont
			{Spielman}},\ }\href {\doibase 10.1038/nature11841} {\bibfield  {journal}
		{\bibinfo  {journal} {Nature}\ }\textbf {\bibinfo {volume} {494}},\ \bibinfo
		{pages} {49} (\bibinfo {year} {2013})}\BibitemShut {NoStop}%
	\bibitem [{\citenamefont {Al-Hilli}\ and\ \citenamefont
		{Evans}(1972)}]{Alhili:jcg15}%
	\BibitemOpen
	\bibfield  {author} {\bibinfo {author} {\bibfnamefont {A.}~\bibnamefont
			{Al-Hilli}}\ and\ \bibinfo {author} {\bibfnamefont {B.}~\bibnamefont
			{Evans}},\ }\href {\doibase http://dx.doi.org/10.1016/0022-0248(72)90129-7}
	{\bibfield  {journal} {\bibinfo  {journal} {J. Cryst. Growth}\ }\textbf
		{\bibinfo {volume} {15}},\ \bibinfo {pages} {93 } (\bibinfo {year}
		{1972})}\BibitemShut {NoStop}%
	\bibitem [{\citenamefont {Dresselhaus}(1955)}]{PhysRev.100.580}%
	\BibitemOpen
	\bibfield  {author} {\bibinfo {author} {\bibfnamefont {G.}~\bibnamefont
			{Dresselhaus}},\ }\href {\doibase 10.1103/PhysRev.100.580} {\bibfield
		{journal} {\bibinfo  {journal} {Phys. Rev.}\ }\textbf {\bibinfo {volume}
			{100}},\ \bibinfo {pages} {580} (\bibinfo {year} {1955})}\BibitemShut
	{NoStop}%
	\bibitem [{\citenamefont {Rashba}(1960)}]{rashba1960properties}%
	\BibitemOpen
	\bibfield  {author} {\bibinfo {author} {\bibfnamefont {E.~I.}\ \bibnamefont
			{Rashba}},\ }\href@noop {} {\bibfield  {journal} {\bibinfo  {journal} {Sov.
				Phys. Solid State}\ }\textbf {\bibinfo {volume} {2}},\ \bibinfo {pages}
		{1224} (\bibinfo {year} {1960})}\BibitemShut {NoStop}%
	\bibitem [{\citenamefont {Kane}\ and\ \citenamefont
		{Mele}(2005)}]{KaneMele:prl2005}%
	\BibitemOpen
	\bibfield  {author} {\bibinfo {author} {\bibfnamefont {C.~L.}\ \bibnamefont
			{Kane}}\ and\ \bibinfo {author} {\bibfnamefont {E.~J.}\ \bibnamefont
			{Mele}},\ }\href {\doibase 10.1103/PhysRevLett.95.226801} {\bibfield
		{journal} {\bibinfo  {journal} {Phys. Rev. Lett.}\ }\textbf {\bibinfo
			{volume} {95}},\ \bibinfo {pages} {226801} (\bibinfo {year}
		{2005})}\BibitemShut {NoStop}%
	\bibitem [{\citenamefont {Bychkov}\ and\ \citenamefont
		{Rashba}(1984)}]{bychkov1984oscillatory}%
	\BibitemOpen
	\bibfield  {author} {\bibinfo {author} {\bibfnamefont {Y.~A.}\ \bibnamefont
			{Bychkov}}\ and\ \bibinfo {author} {\bibfnamefont {E.~I.}\ \bibnamefont
			{Rashba}},\ }\href {http://stacks.iop.org/0022-3719/17/i=33/a=015} {\bibfield
		{journal} {\bibinfo  {journal} {Journal of Physics C: Solid State Physics}\
		}\textbf {\bibinfo {volume} {17}},\ \bibinfo {pages} {6039} (\bibinfo {year}
		{1984})}\BibitemShut {NoStop}%
	\bibitem [{\citenamefont {Manchon}\ \emph {et~al.}(2015)\citenamefont
		{Manchon}, \citenamefont {Koo}, \citenamefont {Nitta}, \citenamefont
		{Frolov},\ and\ \citenamefont {Duine}}]{NewperspectivesRSOC}%
	\BibitemOpen
	\bibfield  {author} {\bibinfo {author} {\bibfnamefont {A.}~\bibnamefont
			{Manchon}}, \bibinfo {author} {\bibfnamefont {H.~C.}\ \bibnamefont {Koo}},
		\bibinfo {author} {\bibfnamefont {J.}~\bibnamefont {Nitta}}, \bibinfo
		{author} {\bibfnamefont {S.~M.}\ \bibnamefont {Frolov}}, \ and\ \bibinfo
		{author} {\bibfnamefont {R.~A.}\ \bibnamefont {Duine}},\ }\href {\doibase
		10.1038/nmat4360} {\bibfield  {journal} {\bibinfo  {journal} {Nature}\
		}\textbf {\bibinfo {volume} {14}},\ \bibinfo {pages} {871} (\bibinfo {year}
		{2015})}\BibitemShut {NoStop}%
	\bibitem [{\citenamefont {Tse}\ and\ \citenamefont
		{Das~Sarma}(2006)}]{PhysRevB.74.245309}%
	\BibitemOpen
	\bibfield  {author} {\bibinfo {author} {\bibfnamefont {W.-K.}\ \bibnamefont
			{Tse}}\ and\ \bibinfo {author} {\bibfnamefont {S.}~\bibnamefont
			{Das~Sarma}},\ }\href {\doibase 10.1103/PhysRevB.74.245309} {\bibfield
		{journal} {\bibinfo  {journal} {Phys. Rev. B}\ }\textbf {\bibinfo {volume}
			{74}},\ \bibinfo {pages} {245309} (\bibinfo {year} {2006})}\BibitemShut
	{NoStop}%
	\bibitem [{\citenamefont {Yuan}\ \emph {et~al.}(2005)\citenamefont {Yuan},
		\citenamefont {Xu}, \citenamefont {Zeng},\ and\ \citenamefont
		{Lu}}]{PhysRevB.72.033320}%
	\BibitemOpen
	\bibfield  {author} {\bibinfo {author} {\bibfnamefont {D.~W.}\ \bibnamefont
			{Yuan}}, \bibinfo {author} {\bibfnamefont {W.}~\bibnamefont {Xu}}, \bibinfo
		{author} {\bibfnamefont {Z.}~\bibnamefont {Zeng}}, \ and\ \bibinfo {author}
		{\bibfnamefont {F.}~\bibnamefont {Lu}},\ }\href {\doibase
		10.1103/PhysRevB.72.033320} {\bibfield  {journal} {\bibinfo  {journal} {Phys.
				Rev. B}\ }\textbf {\bibinfo {volume} {72}},\ \bibinfo {pages} {033320}
		(\bibinfo {year} {2005})}\BibitemShut {NoStop}%
	\bibitem [{\citenamefont {Yang}\ \emph {et~al.}(2006)\citenamefont {Yang},
		\citenamefont {Xu}, \citenamefont {Zeng}, \citenamefont {Lu},\ and\
		\citenamefont {Zhang}}]{PhysRevB.74.075321}%
	\BibitemOpen
	\bibfield  {author} {\bibinfo {author} {\bibfnamefont {C.~H.}\ \bibnamefont
			{Yang}}, \bibinfo {author} {\bibfnamefont {W.}~\bibnamefont {Xu}}, \bibinfo
		{author} {\bibfnamefont {Z.}~\bibnamefont {Zeng}}, \bibinfo {author}
		{\bibfnamefont {F.}~\bibnamefont {Lu}}, \ and\ \bibinfo {author}
		{\bibfnamefont {C.}~\bibnamefont {Zhang}},\ }\href {\doibase
		10.1103/PhysRevB.74.075321} {\bibfield  {journal} {\bibinfo  {journal} {Phys.
				Rev. B}\ }\textbf {\bibinfo {volume} {74}},\ \bibinfo {pages} {075321}
		(\bibinfo {year} {2006})}\BibitemShut {NoStop}%
	\bibitem [{\citenamefont {Ang}\ \emph {et~al.}(2014)\citenamefont {Ang},
		\citenamefont {Cao},\ and\ \citenamefont {Zhang}}]{Ang2014}%
	\BibitemOpen
	\bibfield  {author} {\bibinfo {author} {\bibfnamefont {Y.~S.}\ \bibnamefont
			{Ang}}, \bibinfo {author} {\bibfnamefont {J.~C.}\ \bibnamefont {Cao}}, \ and\
		\bibinfo {author} {\bibfnamefont {C.}~\bibnamefont {Zhang}},\ }\href
	{\doibase 10.1140/epjb/e2014-41015-8} {\bibfield  {journal} {\bibinfo
			{journal} {The European Physical Journal B}\ }\textbf {\bibinfo {volume}
			{87}},\ \bibinfo {pages} {28} (\bibinfo {year} {2014})}\BibitemShut {NoStop}%
	\bibitem [{\citenamefont {Xiao}\ \emph {et~al.}(2016)\citenamefont {Xiao},
		\citenamefont {Xu}, \citenamefont {Van~Duppen},\ and\ \citenamefont
		{Peeters}}]{PhysRevB.94.155432}%
	\BibitemOpen
	\bibfield  {author} {\bibinfo {author} {\bibfnamefont {Y.~M.}\ \bibnamefont
			{Xiao}}, \bibinfo {author} {\bibfnamefont {W.}~\bibnamefont {Xu}}, \bibinfo
		{author} {\bibfnamefont {B.}~\bibnamefont {Van~Duppen}}, \ and\ \bibinfo
		{author} {\bibfnamefont {F.~M.}\ \bibnamefont {Peeters}},\ }\href {\doibase
		10.1103/PhysRevB.94.155432} {\bibfield  {journal} {\bibinfo  {journal} {Phys.
				Rev. B}\ }\textbf {\bibinfo {volume} {94}},\ \bibinfo {pages} {155432}
		(\bibinfo {year} {2016})}\BibitemShut {NoStop}%
	\bibitem [{\citenamefont {Li}\ and\ \citenamefont
		{Carbotte}(2012)}]{PhysRevB.86.205425}%
	\BibitemOpen
	\bibfield  {author} {\bibinfo {author} {\bibfnamefont {Z.}~\bibnamefont
			{Li}}\ and\ \bibinfo {author} {\bibfnamefont {J.~P.}\ \bibnamefont
			{Carbotte}},\ }\href {\doibase 10.1103/PhysRevB.86.205425} {\bibfield
		{journal} {\bibinfo  {journal} {Phys. Rev. B}\ }\textbf {\bibinfo {volume}
			{86}},\ \bibinfo {pages} {205425} (\bibinfo {year} {2012})}\BibitemShut
	{NoStop}%
	\bibitem [{\citenamefont {Mak}\ \emph {et~al.}(2010)\citenamefont {Mak},
		\citenamefont {Lee}, \citenamefont {Hone}, \citenamefont {Shan},\ and\
		\citenamefont {Heinz}}]{PhysRevLett.105.136805}%
	\BibitemOpen
	\bibfield  {author} {\bibinfo {author} {\bibfnamefont {K.~F.}\ \bibnamefont
			{Mak}}, \bibinfo {author} {\bibfnamefont {C.}~\bibnamefont {Lee}}, \bibinfo
		{author} {\bibfnamefont {J.}~\bibnamefont {Hone}}, \bibinfo {author}
		{\bibfnamefont {J.}~\bibnamefont {Shan}}, \ and\ \bibinfo {author}
		{\bibfnamefont {T.~F.}\ \bibnamefont {Heinz}},\ }\href {\doibase
		10.1103/PhysRevLett.105.136805} {\bibfield  {journal} {\bibinfo  {journal}
			{Phys. Rev. Lett.}\ }\textbf {\bibinfo {volume} {105}},\ \bibinfo {pages}
		{136805} (\bibinfo {year} {2010})}\BibitemShut {NoStop}%
	\bibitem [{\citenamefont {Yudin}\ and\ \citenamefont
		{Shelykh}(2016)}]{PhysRevB.94.161404}%
	\BibitemOpen
	\bibfield  {author} {\bibinfo {author} {\bibfnamefont {D.}~\bibnamefont
			{Yudin}}\ and\ \bibinfo {author} {\bibfnamefont {I.~A.}\ \bibnamefont
			{Shelykh}},\ }\href {\doibase 10.1103/PhysRevB.94.161404} {\bibfield
		{journal} {\bibinfo  {journal} {Phys. Rev. B}\ }\textbf {\bibinfo {volume}
			{94}},\ \bibinfo {pages} {161404} (\bibinfo {year} {2016})}\BibitemShut
	{NoStop}%
	\bibitem [{\citenamefont {Kurpas}\ \emph {et~al.}(2016)\citenamefont {Kurpas},
		\citenamefont {Gmitra},\ and\ \citenamefont {Fabian}}]{PhysRevB.94.155423}%
	\BibitemOpen
	\bibfield  {author} {\bibinfo {author} {\bibfnamefont {M.}~\bibnamefont
			{Kurpas}}, \bibinfo {author} {\bibfnamefont {M.}~\bibnamefont {Gmitra}}, \
		and\ \bibinfo {author} {\bibfnamefont {J.}~\bibnamefont {Fabian}},\ }\href
	{\doibase 10.1103/PhysRevB.94.155423} {\bibfield  {journal} {\bibinfo
			{journal} {Phys. Rev. B}\ }\textbf {\bibinfo {volume} {94}},\ \bibinfo
		{pages} {155423} (\bibinfo {year} {2016})}\BibitemShut {NoStop}%
	\bibitem [{\citenamefont {Popovi\ifmmode~\acute{c}\else \'{c}\fi{}}\ \emph
		{et~al.}(2015)\citenamefont {Popovi\ifmmode~\acute{c}\else \'{c}\fi{}},
		\citenamefont {Kurdestany},\ and\ \citenamefont
		{Satpathy}}]{PhysRevB.92.035135}%
	\BibitemOpen
	\bibfield  {author} {\bibinfo {author} {\bibfnamefont {Z.~S.}\ \bibnamefont
			{Popovi\ifmmode~\acute{c}\else \'{c}\fi{}}}, \bibinfo {author} {\bibfnamefont
			{J.~M.}\ \bibnamefont {Kurdestany}}, \ and\ \bibinfo {author} {\bibfnamefont
			{S.}~\bibnamefont {Satpathy}},\ }\href {\doibase 10.1103/PhysRevB.92.035135}
	{\bibfield  {journal} {\bibinfo  {journal} {Phys. Rev. B}\ }\textbf {\bibinfo
			{volume} {92}},\ \bibinfo {pages} {035135} (\bibinfo {year}
		{2015})}\BibitemShut {NoStop}%
	\bibitem [{\citenamefont {Zhou}\ \emph {et~al.}(2015)\citenamefont {Zhou},
		\citenamefont {Shan},\ and\ \citenamefont {Xiao}}]{PhysRevB.91.241302}%
	\BibitemOpen
	\bibfield  {author} {\bibinfo {author} {\bibfnamefont {J.}~\bibnamefont
			{Zhou}}, \bibinfo {author} {\bibfnamefont {W.-Y.}\ \bibnamefont {Shan}}, \
		and\ \bibinfo {author} {\bibfnamefont {D.}~\bibnamefont {Xiao}},\ }\href
	{\doibase 10.1103/PhysRevB.91.241302} {\bibfield  {journal} {\bibinfo
			{journal} {Phys. Rev. B}\ }\textbf {\bibinfo {volume} {91}},\ \bibinfo
		{pages} {241302} (\bibinfo {year} {2015})}\BibitemShut {NoStop}%
	\bibitem [{\citenamefont {Mawrie}\ and\ \citenamefont
		{Ghosh}(2016)}]{0953-8984-28-42-425302}%
	\BibitemOpen
	\bibfield  {author} {\bibinfo {author} {\bibfnamefont {A.}~\bibnamefont
			{Mawrie}}\ and\ \bibinfo {author} {\bibfnamefont {T.~K.}\ \bibnamefont
			{Ghosh}},\ }\href {http://stacks.iop.org/0953-8984/28/i=42/a=425302}
	{\bibfield  {journal} {\bibinfo  {journal} {Journal of Physics: Condensed
				Matter}\ }\textbf {\bibinfo {volume} {28}},\ \bibinfo {pages} {425302}
		(\bibinfo {year} {2016})}\BibitemShut {NoStop}%
	\bibitem [{\citenamefont {Yudin}\ \emph {et~al.}(2017)\citenamefont {Yudin},
		\citenamefont {Gulevich},\ and\ \citenamefont
		{Shelykh}}]{PhysRevB.95.035401}%
	\BibitemOpen
	\bibfield  {author} {\bibinfo {author} {\bibfnamefont {D.}~\bibnamefont
			{Yudin}}, \bibinfo {author} {\bibfnamefont {D.~R.}\ \bibnamefont {Gulevich}},
		\ and\ \bibinfo {author} {\bibfnamefont {I.~A.}\ \bibnamefont {Shelykh}},\
	}\href {\doibase 10.1103/PhysRevB.95.035401} {\bibfield  {journal} {\bibinfo
		{journal} {Phys. Rev. B}\ }\textbf {\bibinfo {volume} {95}},\ \bibinfo
	{pages} {035401} (\bibinfo {year} {2017})}\BibitemShut {NoStop}%
\bibitem [{\citenamefont {Liu}\ and\ \citenamefont {Chang}(2015)}]{liu:nano15}%
\BibitemOpen
\bibfield  {author} {\bibinfo {author} {\bibfnamefont {T.-H.}\ \bibnamefont
		{Liu}}\ and\ \bibinfo {author} {\bibfnamefont {C.-C.}\ \bibnamefont
		{Chang}},\ }\href {\doibase 10.1039/C4NR01600A} {\bibfield  {journal}
	{\bibinfo  {journal} {Nanoscale}\ }\textbf {\bibinfo {volume} {7}},\ \bibinfo
	{pages} {10648} (\bibinfo {year} {2015})}\BibitemShut {NoStop}%
\bibitem [{\citenamefont {Xia}\ \emph {et~al.}(2014)\citenamefont {Xia},
	\citenamefont {Wang},\ and\ \citenamefont {Jia}}]{Xia:nat13}%
\BibitemOpen
\bibfield  {author} {\bibinfo {author} {\bibfnamefont {F.}~\bibnamefont
		{Xia}}, \bibinfo {author} {\bibfnamefont {H.}~\bibnamefont {Wang}}, \ and\
	\bibinfo {author} {\bibfnamefont {Y.}~\bibnamefont {Jia}},\ }\href {\doibase
	10.1038/ncomms5458} {\bibfield  {journal} {\bibinfo  {journal} {Nat.
			Commun.}\ }\textbf {\bibinfo {volume} {5}},\ \bibinfo {pages} {4458}
	(\bibinfo {year} {2014})}\BibitemShut {NoStop}%
\bibitem [{\citenamefont {Rodin}\ \emph {et~al.}(2014)\citenamefont {Rodin},
	\citenamefont {Carvalho},\ and\ \citenamefont {Castro~Neto}}]{Rodin:prl2014}%
\BibitemOpen
\bibfield  {author} {\bibinfo {author} {\bibfnamefont {A.~S.}\ \bibnamefont
		{Rodin}}, \bibinfo {author} {\bibfnamefont {A.}~\bibnamefont {Carvalho}}, \
	and\ \bibinfo {author} {\bibfnamefont {A.~H.}\ \bibnamefont {Castro~Neto}},\
}\href {\doibase 10.1103/PhysRevLett.112.176801} {\bibfield  {journal}
{\bibinfo  {journal} {Phys. Rev. Lett.}\ }\textbf {\bibinfo {volume} {112}},\
\bibinfo {pages} {176801} (\bibinfo {year} {2014})}\BibitemShut {NoStop}%
\bibitem [{\citenamefont {Saberi-Pouya}\ \emph {et~al.}(2016)\citenamefont
	{Saberi-Pouya}, \citenamefont {Vazifehshenas}, \citenamefont {Farmanbar},\
	and\ \citenamefont {Salavati-fard}}]{Samira:drag2016}%
\BibitemOpen
\bibfield  {author} {\bibinfo {author} {\bibfnamefont {S.}~\bibnamefont
		{Saberi-Pouya}}, \bibinfo {author} {\bibfnamefont {T.}~\bibnamefont
		{Vazifehshenas}}, \bibinfo {author} {\bibfnamefont {M.}~\bibnamefont
		{Farmanbar}}, \ and\ \bibinfo {author} {\bibfnamefont {T.}~\bibnamefont
		{Salavati-fard}},\ }\href {http://stacks.iop.org/0953-8984/28/i=28/a=285301}
{\bibfield  {journal} {\bibinfo  {journal} {Journal of Physics: Condensed
			Matter}\ }\textbf {\bibinfo {volume} {28}},\ \bibinfo {pages} {285301}
	(\bibinfo {year} {2016})}\BibitemShut {NoStop}%
\bibitem [{\citenamefont {Magarill}\ \emph {et~al.}(2001)\citenamefont
	{Magarill}, \citenamefont {Chaplik},\ and\ \citenamefont
	{{\'E}ntin}}]{magarill2001spin}%
\BibitemOpen
\bibfield  {author} {\bibinfo {author} {\bibfnamefont {L.}~\bibnamefont
		{Magarill}}, \bibinfo {author} {\bibfnamefont {A.}~\bibnamefont {Chaplik}}, \
	and\ \bibinfo {author} {\bibfnamefont {M.}~\bibnamefont {{\'E}ntin}},\
}\href@noop {} {\bibfield  {journal} {\bibinfo  {journal} {Journal of
		Experimental and Theoretical Physics}\ }\textbf {\bibinfo {volume} {92}},\
\bibinfo {pages} {153} (\bibinfo {year} {2001})}\BibitemShut {NoStop}%
\bibitem [{\citenamefont {B\'acsi}\ and\ \citenamefont
	{Virosztek}(2013)}]{PhysRevB.87.125425}%
\BibitemOpen
\bibfield  {author} {\bibinfo {author} {\bibfnamefont {A.}~\bibnamefont
		{B\'acsi}}\ and\ \bibinfo {author} {\bibfnamefont {A.}~\bibnamefont
		{Virosztek}},\ }\href {\doibase 10.1103/PhysRevB.87.125425} {\bibfield
	{journal} {\bibinfo  {journal} {Phys. Rev. B}\ }\textbf {\bibinfo {volume}
		{87}},\ \bibinfo {pages} {125425} (\bibinfo {year} {2013})}\BibitemShut
{NoStop}%
\bibitem [{\citenamefont {Low}\ \emph {et~al.}(2014{\natexlab{a}})\citenamefont
	{Low}, \citenamefont {Rold\'an}, \citenamefont {Wang}, \citenamefont {Xia},
	\citenamefont {Avouris}, \citenamefont {Moreno},\ and\ \citenamefont
	{Guinea}}]{Low:prl14}%
\BibitemOpen
\bibfield  {author} {\bibinfo {author} {\bibfnamefont {T.}~\bibnamefont
		{Low}}, \bibinfo {author} {\bibfnamefont {R.}~\bibnamefont {Rold\'an}},
	\bibinfo {author} {\bibfnamefont {H.}~\bibnamefont {Wang}}, \bibinfo {author}
	{\bibfnamefont {F.}~\bibnamefont {Xia}}, \bibinfo {author} {\bibfnamefont
		{P.}~\bibnamefont {Avouris}}, \bibinfo {author} {\bibfnamefont {L.~M.}\
		\bibnamefont {Moreno}}, \ and\ \bibinfo {author} {\bibfnamefont
		{F.}~\bibnamefont {Guinea}},\ }\href {\doibase
	10.1103/PhysRevLett.113.106802} {\bibfield  {journal} {\bibinfo  {journal}
		{Phys. Rev. Lett.}\ }\textbf {\bibinfo {volume} {113}},\ \bibinfo {pages}
	{106802} (\bibinfo {year} {2014}{\natexlab{a}})}\BibitemShut {NoStop}%
\bibitem [{\citenamefont {Low}\ \emph {et~al.}(2014{\natexlab{b}})\citenamefont
	{Low}, \citenamefont {Rodin}, \citenamefont {Carvalho}, \citenamefont
	{Jiang}, \citenamefont {Wang}, \citenamefont {Xia},\ and\ \citenamefont
	{Castro~Neto}}]{PhysRevB.90.075434}%
\BibitemOpen
\bibfield  {author} {\bibinfo {author} {\bibfnamefont {T.}~\bibnamefont
		{Low}}, \bibinfo {author} {\bibfnamefont {A.~S.}\ \bibnamefont {Rodin}},
	\bibinfo {author} {\bibfnamefont {A.}~\bibnamefont {Carvalho}}, \bibinfo
	{author} {\bibfnamefont {Y.}~\bibnamefont {Jiang}}, \bibinfo {author}
	{\bibfnamefont {H.}~\bibnamefont {Wang}}, \bibinfo {author} {\bibfnamefont
		{F.}~\bibnamefont {Xia}}, \ and\ \bibinfo {author} {\bibfnamefont {A.~H.}\
		\bibnamefont {Castro~Neto}},\ }\href {\doibase 10.1103/PhysRevB.90.075434}
{\bibfield  {journal} {\bibinfo  {journal} {Phys. Rev. B}\ }\textbf {\bibinfo
		{volume} {90}},\ \bibinfo {pages} {075434} (\bibinfo {year}
	{2014}{\natexlab{b}})}\BibitemShut {NoStop}%
\bibitem [{\citenamefont {Cadiz}\ \emph {et~al.}(2013)\citenamefont {Cadiz},
	\citenamefont {Paget},\ and\ \citenamefont {Rowe}}]{PhysRevLett.111.246601}%
\BibitemOpen
\bibfield  {author} {\bibinfo {author} {\bibfnamefont {F.}~\bibnamefont
		{Cadiz}}, \bibinfo {author} {\bibfnamefont {D.}~\bibnamefont {Paget}}, \ and\
	\bibinfo {author} {\bibfnamefont {A.~C.~H.}\ \bibnamefont {Rowe}},\ }\href
{\doibase 10.1103/PhysRevLett.111.246601} {\bibfield  {journal} {\bibinfo
		{journal} {Phys. Rev. Lett.}\ }\textbf {\bibinfo {volume} {111}},\ \bibinfo
	{pages} {246601} (\bibinfo {year} {2013})}\BibitemShut {NoStop}%
\bibitem [{\citenamefont {Ebeling}\ \emph {et~al.}(2009)\citenamefont
	{Ebeling}, \citenamefont {Blaschke}, \citenamefont {Redmer}, \citenamefont
	{Reinholz},\ and\ \citenamefont {Röpke}}]{1751-8121-42-21-214033}%
\BibitemOpen
\bibfield  {author} {\bibinfo {author} {\bibfnamefont {W.}~\bibnamefont
		{Ebeling}}, \bibinfo {author} {\bibfnamefont {D.}~\bibnamefont {Blaschke}},
	\bibinfo {author} {\bibfnamefont {R.}~\bibnamefont {Redmer}}, \bibinfo
	{author} {\bibfnamefont {H.}~\bibnamefont {Reinholz}}, \ and\ \bibinfo
	{author} {\bibfnamefont {G.}~\bibnamefont {Röpke}},\ }\href
{http://stacks.iop.org/1751-8121/42/i=21/a=214033} {\bibfield  {journal}
	{\bibinfo  {journal} {Journal of Physics A: Mathematical and Theoretical}\
	}\textbf {\bibinfo {volume} {42}},\ \bibinfo {pages} {214033} (\bibinfo
	{year} {2009})}\BibitemShut {NoStop}%
\bibitem [{\citenamefont {Krenner}\ \emph {et~al.}(2006)\citenamefont
	{Krenner}, \citenamefont {Clark}, \citenamefont {Nakaoka}, \citenamefont
	{Bichler}, \citenamefont {Scheurer}, \citenamefont {Abstreiter},\ and\
	\citenamefont {Finley}}]{PhysRevLett.97.076403}%
\BibitemOpen
\bibfield  {author} {\bibinfo {author} {\bibfnamefont {H.~J.}\ \bibnamefont
		{Krenner}}, \bibinfo {author} {\bibfnamefont {E.~C.}\ \bibnamefont {Clark}},
	\bibinfo {author} {\bibfnamefont {T.}~\bibnamefont {Nakaoka}}, \bibinfo
	{author} {\bibfnamefont {M.}~\bibnamefont {Bichler}}, \bibinfo {author}
	{\bibfnamefont {C.}~\bibnamefont {Scheurer}}, \bibinfo {author}
	{\bibfnamefont {G.}~\bibnamefont {Abstreiter}}, \ and\ \bibinfo {author}
	{\bibfnamefont {J.~J.}\ \bibnamefont {Finley}},\ }\href {\doibase
	10.1103/PhysRevLett.97.076403} {\bibfield  {journal} {\bibinfo  {journal}
		{Phys. Rev. Lett.}\ }\textbf {\bibinfo {volume} {97}},\ \bibinfo {pages}
	{076403} (\bibinfo {year} {2006})}\BibitemShut {NoStop}%
\end{thebibliography}
	%

\end{document}